\documentclass[useAMS,usenatbib]{mn2e}
\usepackage{graphicx}

\newcommand{\HII}{H\,{\sc ii}}

\newcommand{\SII}{[S\,{\sc ii}]}
\newcommand{\Halpha}{H${\alpha}$}
\def\p0{\phantom{0}}
\def\SNR{\mbox{{SNR\,J0529--6653}}}
\def\PSR{\mbox{{PSR\,B0529--66}}}

\title[Study of the LMC \SNR\ near \PSR]
  {Multi-frequency study of the Large Magellanic Cloud Supernova Remnant J0529--6653 near Pulsar B0529--66}

\author[L. M. Bozzetto et al.]
  {L. M.~Bozzetto,$^1$ M. D.~Filipovi\'c,$^1$ E. J.~Crawford,$^1$ F.~Haberl,$^2$ M.~Sasaki,$^3$ 
  \newauthor 
  D. Uro{\v s}evi{\'c},$^{4,5}$ W.~Pietsch,$^2$ J. L.~Payne,$^1$ A.~Y.~De~Horta,$^1$ M. Stupar,$^{6,7}$
  \newauthor 
  N. Tothill,$^1$ J. Dickel,$^8$ Y.-H.~Chu,$^9$ and R.~Gruendl,$^9$\\
  $^1$School of Computing and Mathematics, University of Western Sydney
    \break Locked Bag 1797, Penrith South DC, NSW 1797, Australia\\
  $^2$Max-Planck-Institut f\"{u}r extraterrestrische Physik, Giessenbachstra\ss e, D-85748 Garching, Germany\\
  $^3$Institut f\"ur Astronomie und Astrophysik T\"ubingen, Sand 1, D-72076 T\"ubingen, Germany\\
  $^4$Department of Astronomy, Faculty of Mathematics, University of Belgrade, Studentski trg 16, 11000 Belgrade, Serbia\\
  $^5$Isaac Newton Institute of Chile, Yugoslavia Branch \\
  $^6$Department of Physics, Macquarie University, Sydney, NSW 2109, Australia\\
  $^7$Australian Astronomical Observatory, P.O. Box 296, Epping, NSW 1710, Australia\\
  $^8$Physics and Astronomy Department, University of New Mexico, MSC 07-4220, Albuquerque, NM 87131, USA \\
  $^9$Department of Astronomy, University of Illinois, 1002 West Green Street, Urbana, IL 61801, USA
    }
\date{Released 2011 Xxxxx XX}

\pagerange{\pageref{firstpage}--\pageref{lastpage}} \pubyear{2002}

\def\LaTeX{L\kern-.36em\raise.3ex\hbox{a}\kern-.15em
    T\kern-.1667em\lower.7ex\hbox{E}\kern-.125emX}

\begin{document}

\label{firstpage}

\maketitle

 \begin{abstract}
We report the ATCA and \emph{ROSAT} detection of Supernova Remnant (SNR) J0529--6653 in the Large Magellanic Cloud (LMC) which is positioned in the projected vicinity of the known radio pulsar \PSR. In the radio-continuum frequencies, this LMC object follows a typical SNR structure of a shell morphology with brightened regions in the south-west. It exhibits an almost circular shape of D=33$\times$31~pc (1~pc uncertainty in each direction) and radio spectral index of $\alpha$=--0.68$\pm$0.03 -- typical for mid-age SNRs. We also report detection of polarised regions with a peak value of $\sim$17\%$\pm$7\% at 6~cm. An investigation of \emph{ROSAT} images produced from merged PSPC data reveals the presence of extended X-ray emission coincident with the radio emission of the SNR. In X-rays, the brightest part is in the north-east. We discuss various scenarios in regards to the SNR-PSR association with emphasis on the large age difference, lack of a pulsar trail and no prominent point-like radio or X-ray source.
  \end{abstract}


%%%%%%%%%%%%%%%%%%%%%%%%%%%%%%%%%%%%%%%%%%%%%%%%%%%%%%%%%%%%%%%%%%%%%%%%%%%%%%%%%%%%%%%%%%%%%%%%%%%%%%%%%%%%

\section{Introduction}

The Large Magellanic Cloud (LMC) provides an excellent laboratory to study supernova remnants (SNRs) at a known distance of 50~kpc \citep{2008MNRAS.390.1762D}. The line of sight to the LMC lies well away from the Galactic plane, minimising the obscuration and confusion from the foreground gas, dust and stars.

A distinguishing characteristic of SNRs in radio wavelengths is their predominantly non-thermal continuum emission. Generally, SNRs display a radio spectral index of $\alpha\sim-0.5$ (defined by $S\propto\nu^\alpha$), although $\alpha$ may vary significantly, as there exists a wide variety of types of SNRs in different environments and stages of evolution \citep{1998A&AS..127..119F}. For example, younger remnants can have a spectral index of $\alpha\sim-0.8$, while older SNRs and Pulsar Wind Nebulae (PWN) tend to have flatter radio spectra with $\alpha\sim-0.2$. SNRs have a great impact on the physical properties, structure, and evolution of the interstellar medium (ISM). Conversely, the interstellar environments in which SNRs reside will heavily affect the remnants' evolution.

Type~II supernovae result from the core collapse of massive stars with initial masses greater than \mbox{$\sim 8\pm1M_{\sun}$} \citep{2009ARA&A..47...63S}, and may leave behind compact central objects, such as neutron stars that may be observable as pulsars. However, not many SNRs from Type~II supernovae are observed to host pulsars. Among the 51 confirmed and 25 possible candidate SNRs in the LMC \citep[][Filipovi\'c et al.\ in prep]{2010ApJ...725.2281K, 2010AJ....140..584D}, only three (N\,49, 30~Dor~B, B0540--693) have documented associations with a pulsar \citep{2006ApJ...649..235M}. There are over 70 well studied SNRs in the Magellanic Clouds (MCs) as well as 19 detected pulsars. \citet{2010MNRAS.406L..80R} modelled the potentially observable radio pulsars in the MCs and predicted some 1.79$\times$10$^{4}$ pulsars. The Milky Way (MW) has 274 SNRs \citep{2009BASI...37...45G} and $\sim$1900 pulsars. Therefore, the observed SNR/PSR ratio in the MCs is much less than in the MW ($\sim$1/20). However, if the pulsar surveys are sensitivity limited, then the expected ratio of the mean distances $\sim$(10/50)$^{2}$ would result in the two fractions being similar. This rarity of pulsar-SNR connections may be attributed to the fact that many young neutron stars in SNRs exhibit different properties from the traditional radio pulsars \citep{2000ASPC..202..699G}. Radio pulsars remain detectable well after their SNRs have merged into the ISM; therefore, most pulsars are not in SNRs and their general properties may be different from those young ones in SNRs.

\PSR\ in the LMC was the first extra-galactic pulsar discovered. It was found by \citet{1983Natur.303..307M} using the 64-m radio telescope at the Parkes Observatory. The pulsar was studied by \citet{1991MNRAS.252...13C} at a frequency of 600~MHz with the intent of using a polarimeter to collect integrated profiles, rotation and dispersion measures. \citet{2001ApJ...553..367C} used the Parkes telescope at 20~cm and improved the radio positional accuracy to \mbox{RA~(J2000)=05$^h$29$^m$50.92$^s$($\pm$0.13$^s$)} and \mbox{DEC~(J2000)=$-$66\degr52\arcmin38.2\arcsec($\pm$0.9\arcsec)}. \citet{2005yCat.7245....0M} then included new and more detailed information on the pulsar in a catalogue along with 1532 other pulsars and in 2006 re-observed the pulsar at Parkes at a wavelength of 20~cm. \citet{1998A&AS..130..441F} detected the radio source LMC\,B0530--6655 and suggested it to be an SNR candidate based on its non-thermal spectral index. We note that the \SNR\ (or LMC\,B0530--6655) lies just on the northeast edge of the arc of H$\alpha$ nebulosity DEM~L214 \citep{1976MmRAS..81...89D}. Finally, \citet[][hereafter HP99]{1999A&AS..139..277H} detected a nearby \emph{ROSAT} X-ray source ([HP99]\,440) at a position of \mbox{RA~(J2000)=05$^h$29$^m$59.3$^s$} and \mbox{DEC~(J2000)=$-$66\degr53\arcmin13\arcsec.} 

Here, we report on radio-continuum, X-ray and optical observations of the candidate LMC \SNR\ with its possible association with the LMC pulsar \PSR. Observations, data reduction and imaging techniques are described in Sect.~\ref{section:observations}. The astrophysical interpretation of the moderate-resolution total intensity and polarimetric images are discussed in Sect.~\ref{section:rad}.

%%%%%%%%%%%%%%%%%%%%%%%%%%%%%%%%%%%%%%%%%%%%%%%%%%%%%%%%%%%%%%%%%%%%%%%%%%%%%%%%%%%%%%%%%%%%%%%%%%%%%%%%%%%%

\section{Observations and data reduction}
 \label{section:observations}

\subsection{Radio-continuum}
 \label{datareduction_radio}

\begin{figure}
\includegraphics[angle=-90,scale=0.4]{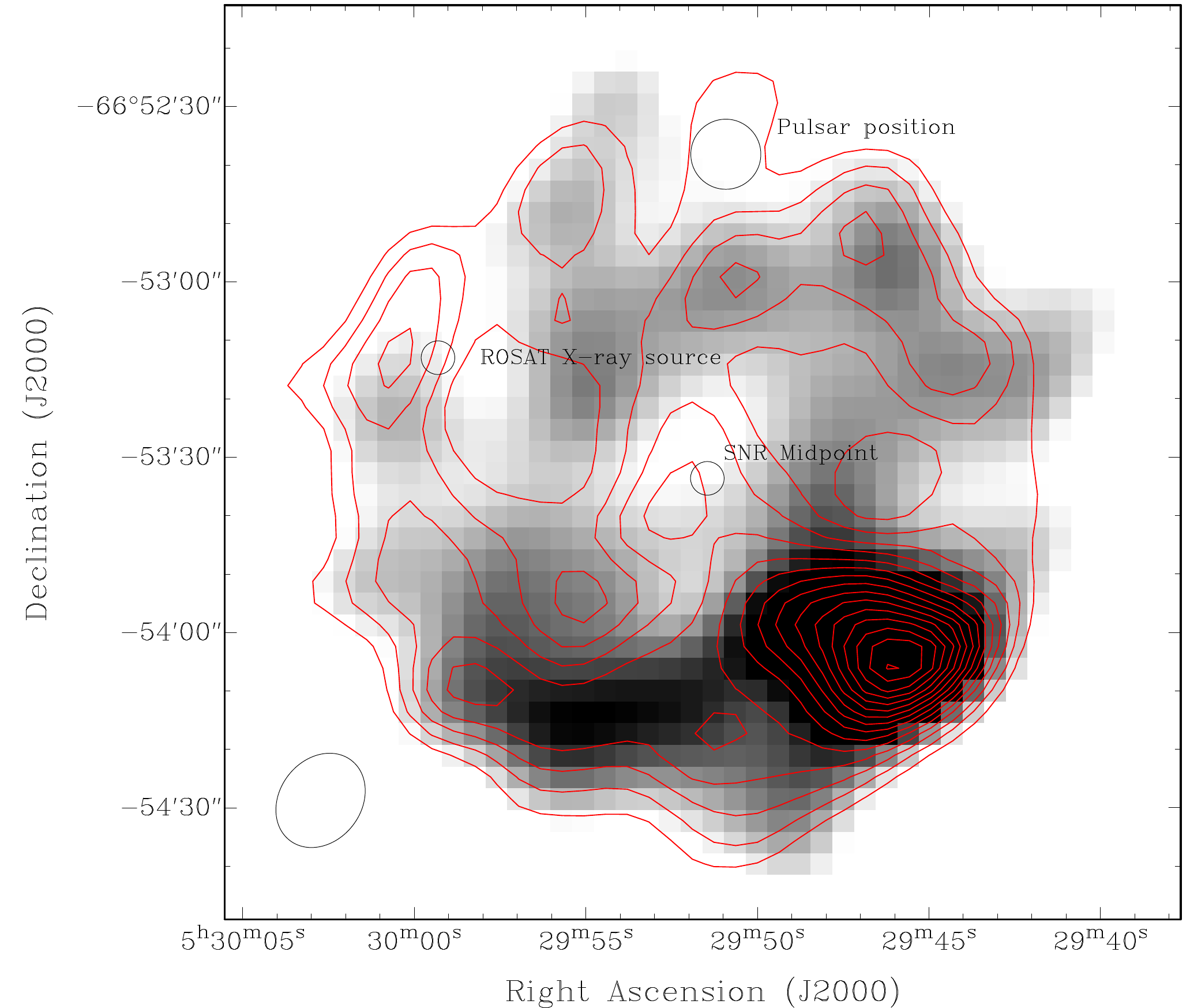}
 \caption{ATCA image of \SNR\ at 3~cm overlaid with 6~cm contours. The contours are from 3 until 17 $\sigma$ with spacings of 1~$\sigma$ (0.11~mJy). The black ellipse in the lower left corner represents the synthesised beam width of 17.3\,\arcsec$\times$14.0\arcsec. The circle in the centre marks the midpoint of the SNR. The circle in the east shows the catalogued X-ray location of [HP99]\,440 and the circle in the north is the position of the pulsar \PSR.}
  \label{fig1}
\end{figure}

We used radio observations at four frequencies (Table~\ref{tbl-1}) to study and measure flux densities of \SNR. For the 36~cm (Molonglo Synthesis Telescope -- MOST) flux density measurement given in Table~\ref{tbl-1} we used unpublished  images as described by \citet{1984AuJPh..37..321M} and for the 20~cm we used image from \citet{2007MNRAS.382..543H}. Two Australia Telescope Compact Array (ATCA) projects (C634 and C797; at 6/3cm) observations were combined with mosaic observations from project C918 \citep{2005AJ....129..790D}. Data for project C634 were taken by the ATCA on 1997 August~2, using the array configuration EW375. Four days of observations were taken from project C797: 1999 May 1--2 (array configuration: 1.5C), 1999 July~21 (array configuration 750D), and 1999 July 31 (array configuration: 1.5D). For the final image (stokes parameter \textit{I}) we exclude baselines created with the $6^\mathrm{th}$ ATCA antenna, leaving the remaining five antennas to be arranged in a compact configuration. C634 observations were carried out in ``snap-shot'' mode, totalling $\sim$1 hour of integration over a 12~hour period. Source PKS~B1934--638 was used as the primary calibrator and source PKS~B0530--727 was used as the secondary calibrator. The \textsc{miriad} \citep{2006Miriad} and \textsc{karma} \citep{2006Karma} software packages were used for reduction and analysis. The 6 and 3~cm images (Fig.~\ref{fig1}) were constructed using \textsc{miriad} multi-frequency synthesis \citep{1994A&AS..108..585S}. Deconvolution was achieved with the {\sc clean} and {\sc restor} tasks with primary beam correction applied using the {\sc linmos} task. Similar procedures were used for the \textit{U} and \textit{Q} stokes parameters. 

\begin{table}
 \caption{Integrated flux densities of \SNR. The flux density at $\lambda$=36~cm was estimated using images from \citet{1984AuJPh..37..321M} and at $\lambda$=20~cm from \citet{2007MNRAS.382..543H}.}
 \label{tbl-1}
 \begin{tabular}{@{}ccccc}
  \hline
  $\nu$ & $\lambda$ & Beam Size        & R.M.S& S$_\mathrm{Total}$  \\
  (MHz) & (cm)      & (\arcsec)        & (mJy)& (mJy)               \\
  \hline
 \p0843 & 36        & 43.0$\times$43.0 & 0.50 & 99                  \\
   1377 & 20        & 40.0$\times$40.0 & 0.50 & 69                     \\
   4800 & \p06      & 17.3$\times$14.0 & 0.11 & 28                     \\
   8640 & \p03      & 17.3$\times$14.0 & 0.06 & 21                   \\
  \hline
 \end{tabular}
\end{table}

\subsection{Optical}
 \label{datareduction_optical}

\begin{figure}
\includegraphics[scale=.83]{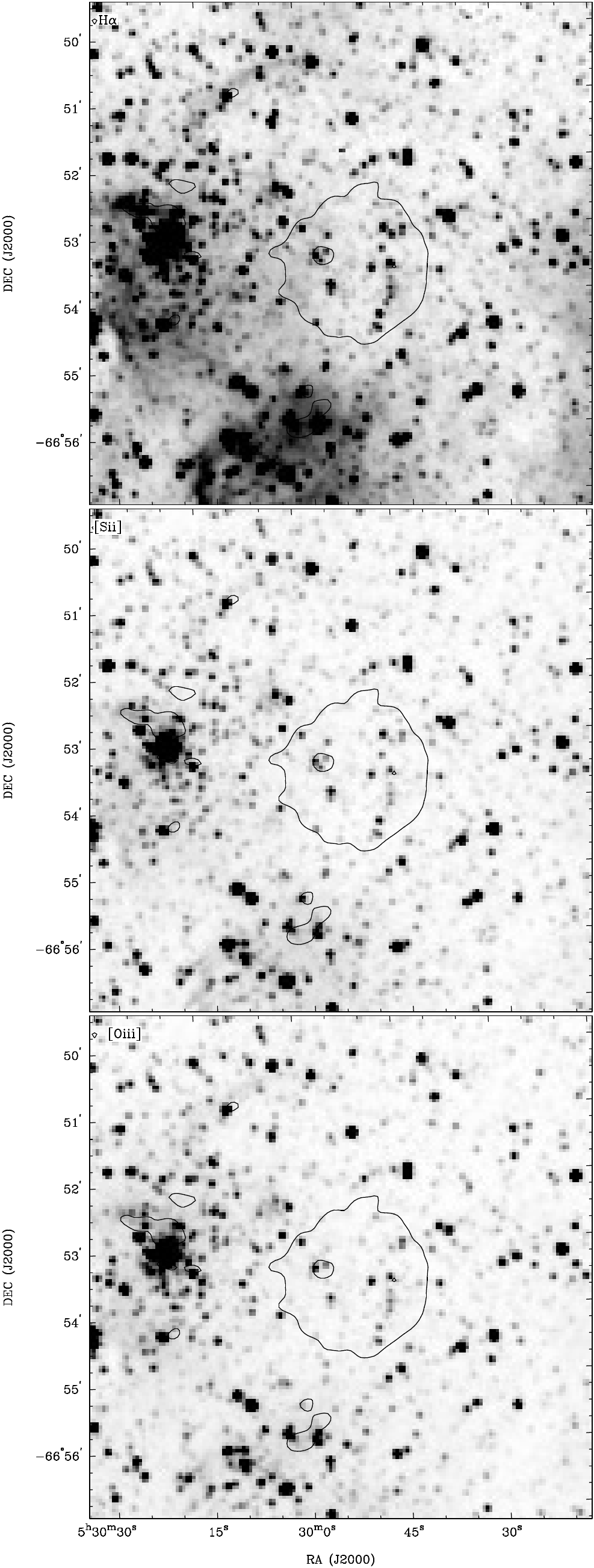}
\caption{MCELS H$\alpha$ ({\it top}), [S\,{\sc ii}] ({\it middle}) and [O\,{\sc iii]} ({\it bottom}) images of the area around \SNR\ overlaid with 6~cm 3$\sigma$ (0.33~mJy) radio-continuum contour. 
 \label{fig2}}
\end{figure}

\begin{figure}
 \begin{center}
\includegraphics[angle=-90,scale=0.435]{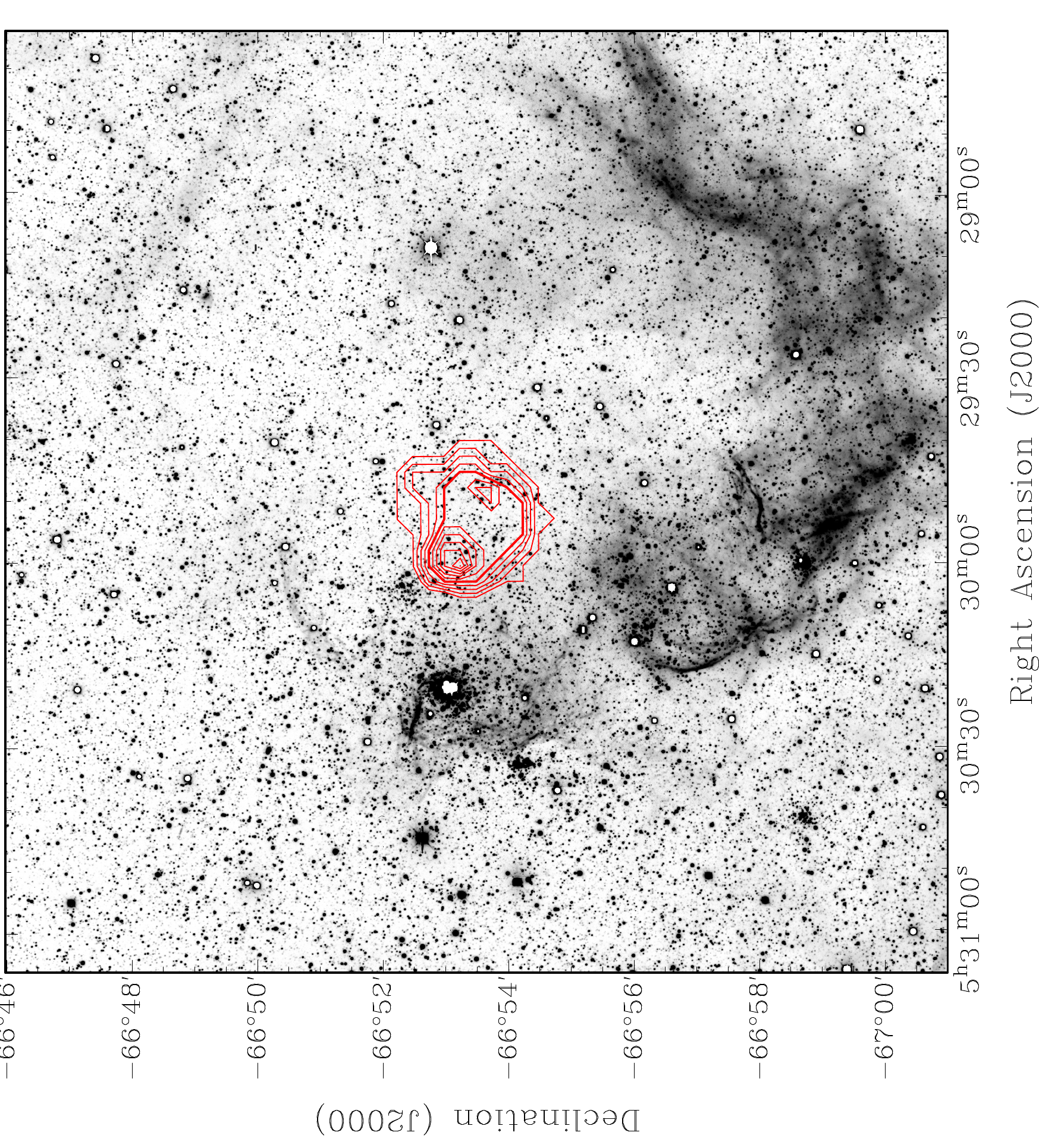}
 \caption{The MCELS-2 H$\alpha$ image of the area \SNR\ overlaid with the \emph{ROSAT} PSPC contours from the broad-band (0.1--2.4~keV) image (pixel size: 15\arcsec). Contours are from 3 to 7~cts/pixel with 0.5~cts/pixel spacings.  }
 \label{fig3}
 \end{center}
\end{figure}

We have examined images (Fig.~\ref{fig2}) from the Magellanic Clouds Emission Line Survey (MCELS) \citep{2006NOAONL.85..6S} and higher-resolution H$\alpha$ images (Fig.~\ref{fig3}) obtained with the MOSAIC~{\sc ii} camera on the Blanco 4-m telescope at the Cerro Tololo Inter-American Observatory. The extended arc of the \HII\ region DEM\,L214 is readily seen but we do not find any optical nebulosity associated with the SNR candidate or the \PSR. Additional long-slit observations were performed on 2010 September 20, using the 1.9-m telescope and Cassegrain spectrograph at the South African Astronomical Observatory (SAAO) in Sutherland. While the observing conditions were good, we did not detect any emission from the SNR candidate.

\subsection{X-rays}
 \label{datareduction_Xray}

\begin{figure}
 \begin{center}
\includegraphics[scale=0.5]{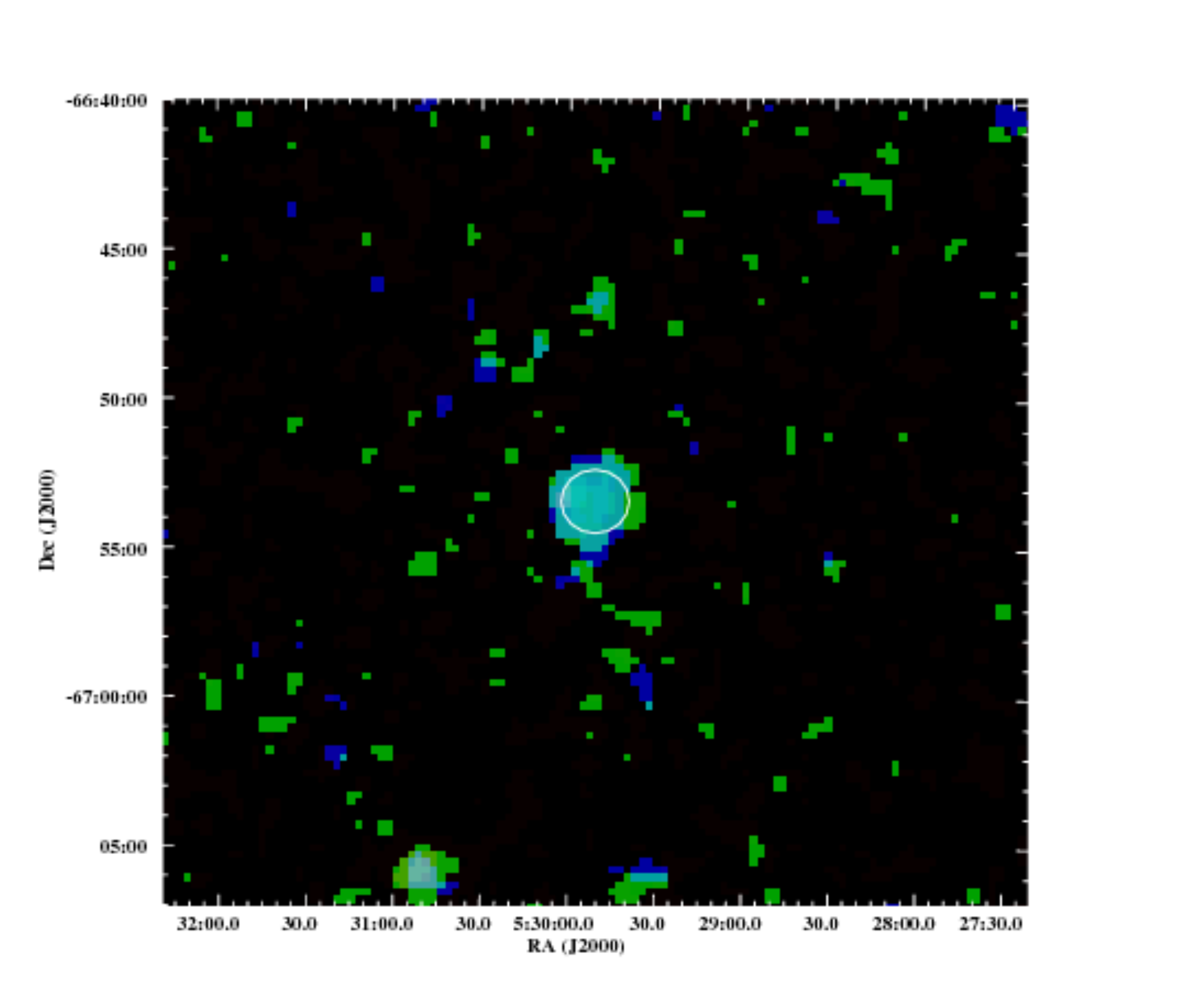}
 \caption{The \emph{ROSAT} PSPC RGB colour image of the area around \SNR. The energy bands are: red (0.1--0.4 keV), green (0.5--0.9~keV) and blue (0.9--2.0~keV). The image has a pixel size of 15\arcsec\ and is smoothed with a $\sigma$ of one pixel. A white ellipse is centred on the position of the \SNR\ with extent of 137\arcsec$\times$128\arcsec\ (33$\times$31~pc). }
 \label{fig4}
 \end{center}
\end{figure}

\emph{ROSAT} performed a raster of pointed observations with the Position Sensitive Proportional Counter (PSPC) to map the soft X-ray emission from the hot gas in the LMC super giant shell-4 \citep[\mbox{SGS~LMC-4}]{1980MNRAS.192..365M,1999IAUS..190..158B}. A number of these observations with typical exposures of 1000\,s include \SNR\ (Figs.~\ref{fig3} and \ref{fig4}). The HP99 catalogue includes a very weak detection within the extent of the radio emission from \SNR\ which does not allow to derive much about the X-ray properties of the source (the position of [HP99]\,440 is shown in Fig.~\ref{fig1}). The hardness ratios are undefined and therefore give no information on the spectrum. The catalogue of HP99 was derived from the individual PSPC observations, not utilising the combined exposure of overlapping fields. To investigate [HP99]\,440 and its possible association with \SNR\ and \PSR\ in more detail we selected 13 observations from the SGS~LMC-4 raster which covered the SNR/PSR within 24\arcmin\ of the optical axis (to avoid the degraded point spread function at larger off-axis angles). We produced images in different energy bands (broad: 0.1--2.4~keV, soft: 0.1--0.4~keV, hard1: 0.5--0.9~keV and hard2: 0.9--2.0~keV) from the merged data. A colour image of the area around the SNR with net exposure (vignetting corrected) of 11.0~ks is shown in Fig.~\ref{fig4} with red, green and blue representing the X-ray intensities in the soft, hard1 and hard2 bands. The resolution of the \emph{ROSAT} PSPC varies with energy but the point spread function is always less than 1\arcmin.

%%%%%%%%%%%%%%%%%%%%%%%%%%%%%%%%%%%%%%%%%%%%%%%%%%%%%%%%%%%%%%%%%%%%%%%%%%%%%%%%%%%%%%%%%%%%%%%%%%%%%%%%%%%%

\section{Results and Discussion}
 \label{section:rad}

\begin{figure*}
\centering\includegraphics[angle=-90,scale=.3075]{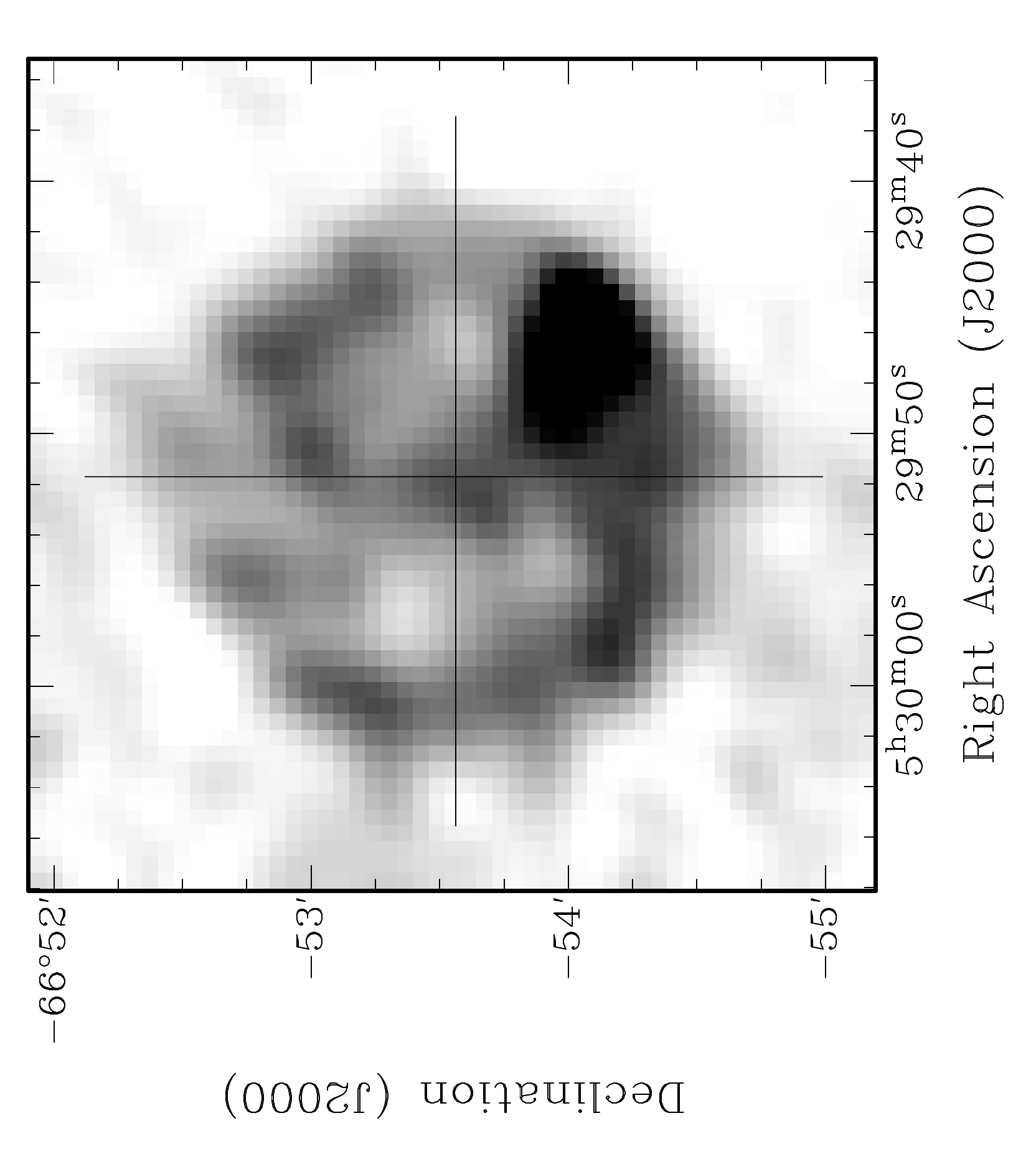}
\centering\includegraphics[angle=-90,scale=.275]{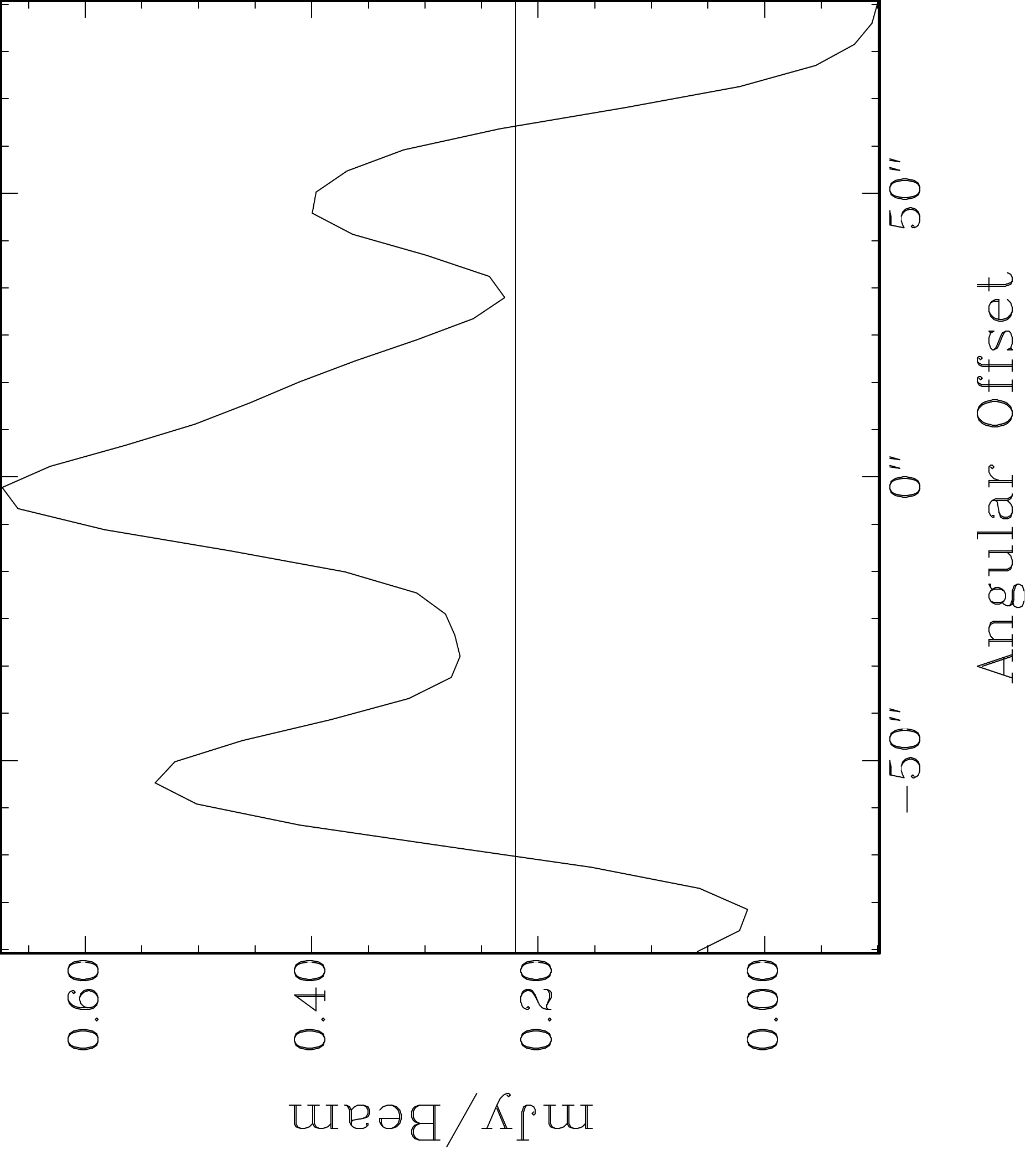}
\centering\includegraphics[angle=-90,scale=.275]{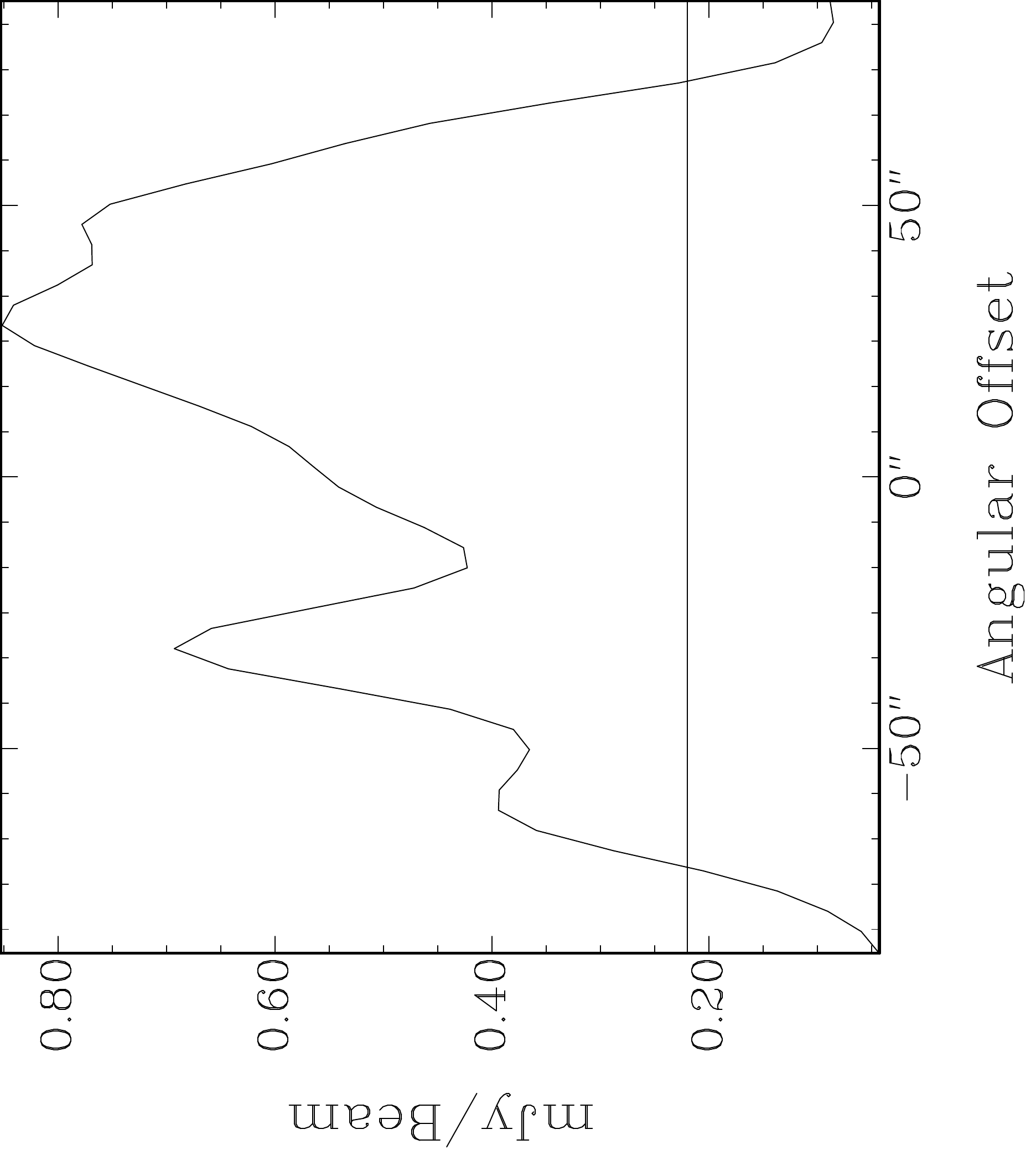}
\caption{The left image shows the 6~cm image overlaid with the major and minor axis labels. The centre and right images show the I-Profile of the major and minor axis respectively, with an overlaid black line at 3$\sigma$.
 \label{fig5}}
\end{figure*}

The radio-continuum remnant \SNR\ exhibits a ring-like morphology, indicative of a shell structure, with brightened region along the southwest rim (Fig.~\ref{fig1}). It is centred at \mbox{RA~(J2000)=05$^h$29$^m$51.0$^s$} and \mbox{DEC~(J2000)=--66\degr53\arcmin27.1\arcsec.} We estimate the spatial extent of \SNR\ (Fig.~\ref{fig5}) at the 3$\sigma$ (Table~\ref{tbl-1}; Col.~4) level (0.33~mJy) along the major (N-S) and minor (E-W) axes. Its size at 6~cm is 137\arcsec$\times$128\arcsec$\pm$4\arcsec\ (33$\times$31~pc with 1~pc uncertainty in each direction). We also estimate the \SNR\ ring thickness to $<$20\arcsec\ (5~pc) at 6~cm, about 30\% of the SNR's radius. 

\begin{figure}
 \begin{center}
\includegraphics[angle=-90,scale=0.41]{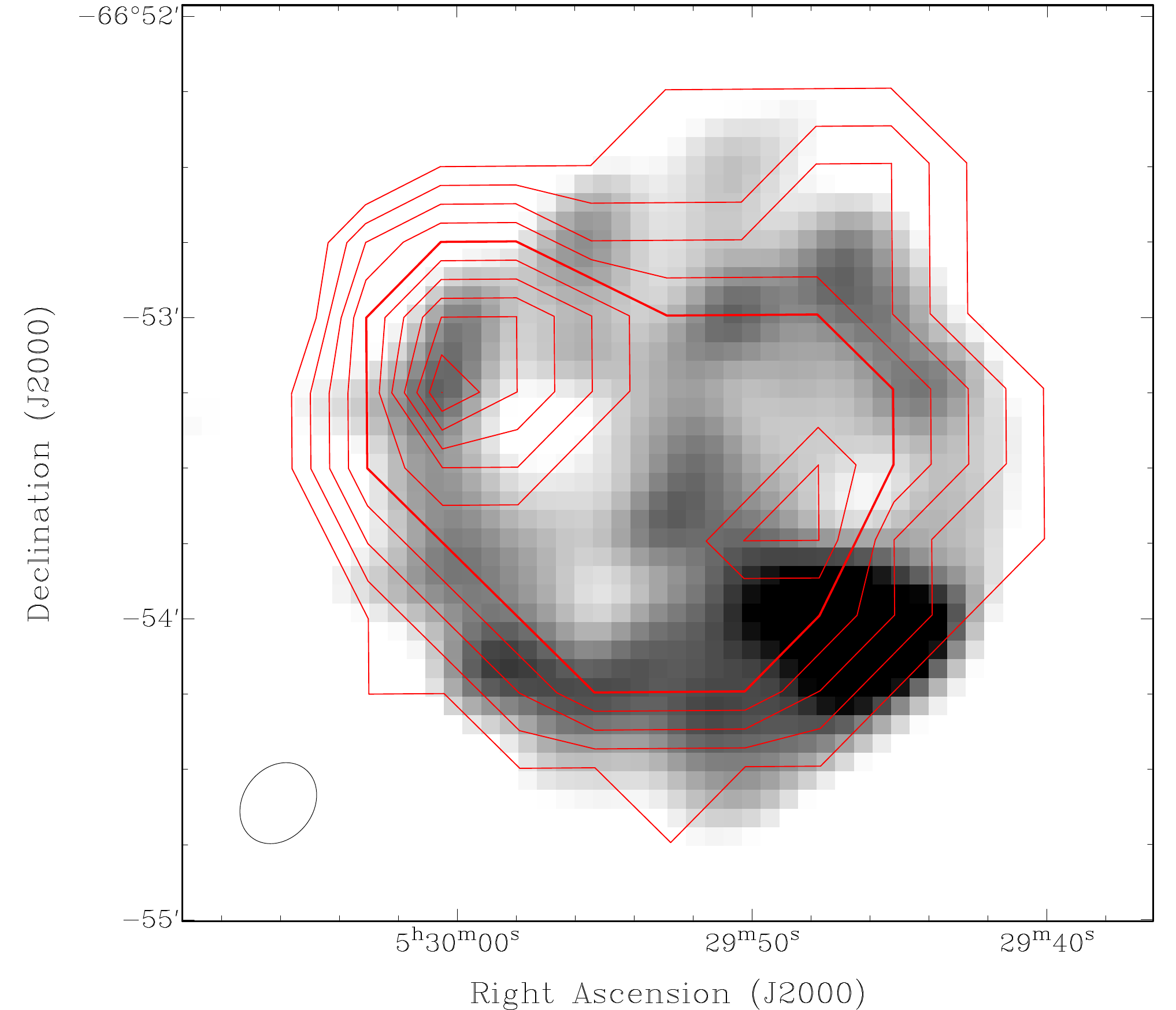}
 \caption{ATCA image of \SNR\ at 6~cm overlaid with the contours from \emph{ROSAT} PSPC image (0.1--2.4~keV). Contours are as in Fig.~\ref{fig3}. The ellipse in the bottom left corner represent the beam of 17.3\arcsec $\times$14.0\arcsec.}
 \label{fig6}
 \end{center}
\end{figure}

An extended X-ray source is clearly seen at the location of the SNR with a brighter spot right at the position of [HP99]\,440. Our images are based on the combination of data from 13 observations, which had too short exposures to allow an astrometric alignment to a common reference system. Therefore, in order to verify the source extent we compared the projected profiles of the extended source and other nearby sources, which appear point-like. For the profile, we projected the counts in the 0.1$-$2.4 keV image within a slice of 220\arcsec\ width and position angle of 330\degr\ (to investigate the profile of the bright spot). We measure a FWHM of 130\arcsec$\pm$7\arcsec\ for the extended source and 75\arcsec$\pm$7\arcsec\ for the source visible at the southern rim of the image in Fig.~\ref{fig4}, which exhibits a similar peak intensity. The extended profile shows a peak on top of the broader intensity distribution confirming the brightest part of the X-ray emission in the north-east. The presence of extended X-ray emission is coincident with the radio emission of the SNR (Fig.~\ref{fig6}). 

The latter is also supported from the relation between X-ray luminosity and energy loss for rotation- powered pulsars (L$_{\rm x}$=(10$^{-4}$--10$^{-2}$)$\times\dot{\rm E}$; \citet{1997A&A...326..682B}). With a value of $\dot{\rm E}$=6.6$\times10^{32}$~erg~s$^{-1}$ (from online ATNF pulsar catalogue\footnote{http://www.atnf.csiro.au/research/pulsar/psrcat/}) for \PSR\ we expect an X-ray luminosity for the pulsar of well below $10^{31}$ erg s$^{-1}$, several orders of magnitude below the detection limit of $\sim10^{34}$~erg~s$^{-1}$ for a 10~ks \emph{ROSAT} PSPC observation. Also, from the estimated spin down age of $\sim$10$^6$~yr (see also below) one would not expect the existence of a PWN.

The net counts of 337$\pm$25~cts~s$^{-1}$ derived from the SNR in the 0.1$-$2.4 keV band are insufficient for spectral modelling in order to derive temperature and density from which one could estimate the explosion energy. Assuming a thermal emission model (APEC) with a temperature kT=0.19~keV, typical for faint SNR in the Magellanic Clouds \citep{2008A&A...485...63F} and a foreground absorption of 6$\times$10$^{20}$~H~cm$^{-2}$ we can roughly estimate the absorption-corrected X-ray luminosity to 5.7$\times$10$^{35}$~erg~s$^{-1}$.

We estimate the radio spectral index $\alpha$=$-$0.68$\pm$0.03 from the integrated flux densities given in Table~1. This spectral index is consistent with those of known SNRs of mid-to-younger ages. \SNR\ is located in the centre of the SGS~LMC-4 (some $\sim$3\degr\ from 30~Dor) and therefore, conceivable that it is expanding in a very low density environment, which causes a slightly steeper spectral index for its age (see Sect.~\ref{section:rad}; Para.~10). 

\begin{figure}
 \includegraphics[width=0.525\textwidth]{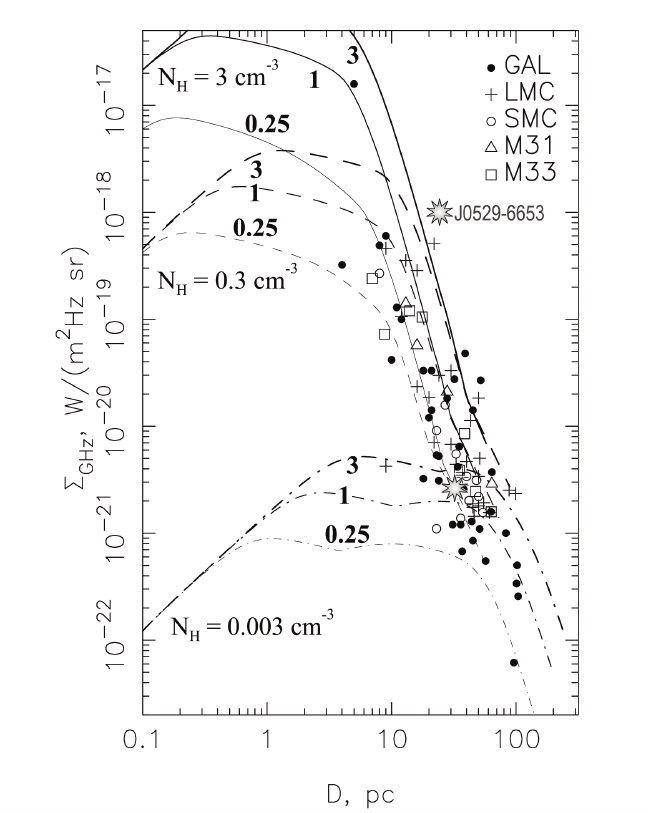}
 \caption{Surface brightness-to-diameter diagram from \citet{2004A&A...427..525B}, with \SNR\ added. The evolutionary tracks are for ISM densities of N$_{H}$= 3, 0.3 and 0.003~cm$^{-3}$ and explosion energies of E$_{SN}$ = 0.25, 1 and 3$\times10^{51}$~erg.}
 \label{fignew}
\end{figure}

Fig.~\ref{fignew} shows a surface brightness--diameter ($\Sigma-D$) diagram at 1~GHz with theoretically-derived evolutionary tracks \citep{2004A&A...427..525B} superposed. \SNR\ lies at $(D,\Sigma)$ = (32~pc, $2.3\times 10^{-21}$~W m$^{-2}$~Hz$^{-1}$~Sr$^{-1}$) on the diagram. Its position tentatively suggests that it is in the early Sedov phase of evolution --- expanding into a very low density environment with the higher initial energy of a supernova explosion ($2-3\times10^{51}$~ergs) and the age of $\sim$25\,000~yr. We acknowledge that such scenario couldn't explain for the lack of SNR's optical emission. Alternatively, this SNR could be bit older ($\sim$70\,000~yr; the end of adiabatic phase) and expanding into the medium dense environment with the minimal start-up energy.

\begin{figure*}
\includegraphics[angle=-90,scale=.6]{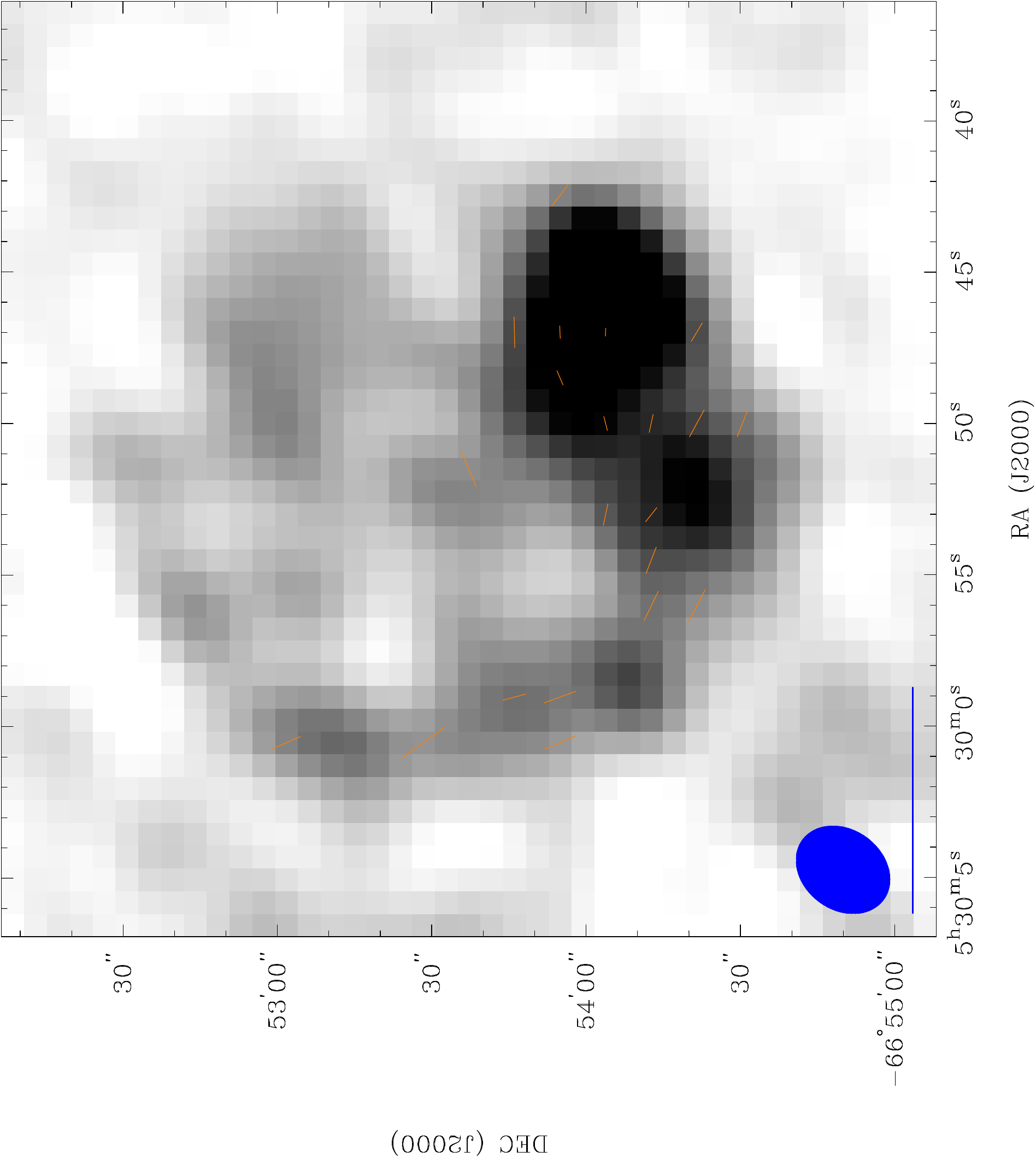}
\caption{\SNR\ at 6~cm overlaid with polarisation ($E$) vectors that peak at 17$\pm$7\%. The ellipse in the bottom left corner represent the beam of 17.3\arcsec $\times$14.0\arcsec\ and the line under it representing a polarisation $E$ vector of 100\%. 
 \label{fig7}}
\end{figure*}

A linear polarisation image was created for 6~cm wavelength using \textit{Q} and \textit{U} parameters (Fig.~\ref{fig7}). The mean fractional polarisation was calculated using flux density and polarisation:
%\[
P=$\frac{\sqrt{S_{Q}^{2}+S_{U}^{2}}}{S_{I}}\cdot 100\%$,
%\]
\noindent where $S_{Q}, S_{U}$ and $S_{I}$ are integrated intensities for the \textit{Q}, \textit{U} and \textit{I} Stokes parameters. Our estimated peak value is 17\%$\pm$7\% (3$\sigma$) at 6~cm and no reliable detection at 3~cm. Along the south side of the SNR shell there is a pocket of uniform polarisation (Fig.~\ref{fig7}), possibly indicating varied dynamics along the shell. The polarisation appear to be uniformly distributed (tangential orientation) coinciding with the SNR total intensity. As the whole SNR is nearly round the only irregularity is the brightness variations. While the distance from the nearby \Halpha\ filament (DEM\,L214) seems bit far away it is most likely that the SNR interacted with some less obvious local clouds so the radio-continuum shape has not yet been disturbed.

Without reliable polarisation measurements at a second frequency (3~cm) we cannot determine the Faraday rotation and thus cannot deduce the magnetic field strength. However, by using the new equipartition formula for SNRs \citep{arbutina}, we can estimate the magnetic field strength for the \SNR. The derivation of the new equipartition formula is based on the \citet{1978MNRAS.182..443B} diffuse shock acceleration (DSA) theory. This derivation is purely analytical, accommodated especially for the estimation of magnetic field strength in SNRs. Using this new formula, the calculated magnetic field strengths for SNRs are between those calculated by using classical equipartition \citep{1970ranp.book.....P} and revised equipartition \citep{2005AN....326..414B}. The average equipartition field over the whole shell of \SNR\ is $\sim$48~$\mu$G (see \citet{arbutina}; and corresponding "calculator"\footnote{The calculator is available on http://poincare.matf.bg.ac.rs/\~{}arbo/eqp/ }), corresponding those of young to middle-aged SNRs where the interstellar magnetic field is compressed and amplified by the strong shocks.

The Parkes pulsar catalogue suggests that the pulsar \PSR\ is 9.97$\pm~0.77\times$10$^{5}$~yr old. SNRs normally merge into the ISM and become unrecognisable after $\sim$150\,000~yr. We point out that a measured radius of $\sim$17~pc would imply a mean expansion speed of only about 0.1~km~s$^{-1}$ for a million year age which is clearly unreasonable. Therefore, this factor of $\sim$6 in age difference would rule out the possibility that the pulsar and SNR were formed from a common supernova explosion. However, \citet{2002ApJ...567L.141M} showed that pulsar age estimates from its ``characteristic age'' \mbox{($\tau_{c}$ = P/2\.{P})} could be uncertain when compared to its corresponding SNR age. Also for nearby Galactic isolated neutron stars large differences between the  characteristic age and the dynamic age estimated from their proper motion and the likely birth place are found \citep[a factor of $\sim$8 for the case of RX\,J1856.5-3754, see][]{2011arXiv1107.1673T}. They also argue that the ratio depends on the magnetic field strength and its evolution in the past.

If we assume that the \SNR\ and pulsar \PSR\ are associated and that the pulsar was born near the geometric centre of the SNR, then the pulsar must have moved $\sim$14$\pm$1~pc to its current location \citep{2001ApJ...553..367C}. Assuming the above ``characteristic pulsar age'', the pulsar would be travelling (in the sky plane) at \mbox{$\sim$~14~km~s$^{-1}$,} an absolute minimum kick velocity. If we assume a canonical SNR age of 25\,000--50\,000~yr then the pulsar velocity in the sky plane would be in a range of 550 to 275~km~s$^{-1}$, in agreement with the typical pulsar kick velocities \citep{1994Natur.369..127L}. The direction of the pulsar's travel is not necessarily in the sky plane, so this is a lower limit. However, unless the direction of the pulsar motion is significantly out of the sky plane, its real velocity will not be much greater. \citet{2002AJ....124.2135K} found a jet-like structure and a point X-ray source within the LMC~SNR~N\,206 travels at a speed of 800~km~s$^{-1}$ and \citet{2011A&A...530A.132O} estimated the kick velocity of the pulsar candidate inside SMC~PWN~IKT\,16 to be approximately 580~km~s$^{-1}$. The possible pulsars have left prominent radio trails in N\,206 and IKT\,16, but no such trail can be seen in our radio images of \SNR. This makes a SNR-pulsar connection less likely. Nevertheless, if \PSR\ is related to the SNR, it is likely to be travelling more slowly and thus any trail would be less prominent. We acknowledge that even if the pulsar is not traveling, there may still be a prominent PWN in the X-rays, although it will not be trailed. Unfortunately, the lack of evidence of PWN in the \emph{ROSAT} data is purely because of resolution, and therefore, cannot be taken as an evidence of lacking a PWN. 

The absence of detectable optical emission (Fig.~\ref{fig2}) from this SNR is not unique as other well-studied SNRs, such as LMC SNR\,J0528--6714 \citep{2010A&A...518A..35C} or the Galactic Vela~Jr. \citep{2005AdSpR..35.1047S}, also do not exhibit optical emission. The ISM in the interior of SGS~LMC-4 must have very low density and hence the lack of optical emission. We argue that, as with the majority of other SNRs in the Magellanic Clouds, this intriguing SNR is most likely in the adiabatic phase of its evolution \citep{2008MNRAS.383.1175P} simply because of its modest size. 

In the MCELS-2 \Halpha\ image (Fig.~\ref{fig3}) we clearly see the large extent of \HII\ region DEM\,L214 in the south, south-west and south-east. We also see smaller scale filamentary structures in the north of the remnant which, together with DEM\,L214 and the \Halpha\ emission around the cluster in the east, seem to form a large scale shell. This and the distribution of the massive stars (see next paragraph) indicate that some heating must have already taken place where the SNR is located. A shock expanding in a already hot thin medium is not efficient and expands quickly. This may explain the faint X-ray appearance of the remnant \SNR.

\begin{figure*}
 \begin{center}
\includegraphics[scale=1.0]{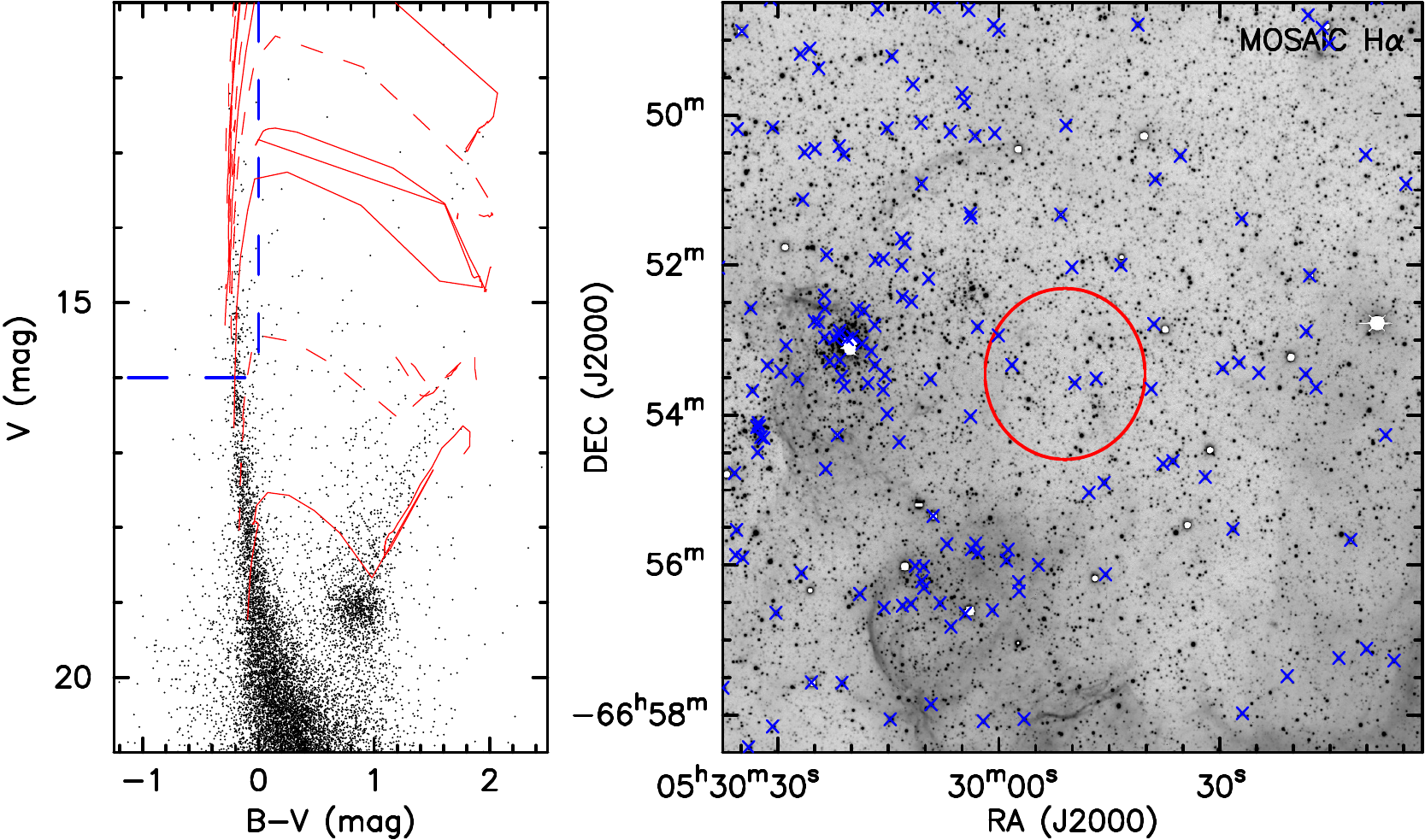}
 \caption{Left panel shows a $B-V$ vs. $V$ colour-magnitude diagram from the MCPS \citep{2004AJ....128.1606Z}. Stellar evolutionary tracks from \citet{2001A&A...366..538L} are shown with solid red lines (5, 15, 25 and 60 $M_\odot$) and dashed red lines (3, 9, 20 and 40 $M_\odot$). The selection criteria used for probable B-star candidates is shown as a heavy dashed blue lines. On the right is the MCELS-2 \Halpha\ image of the area around \SNR\ overlaid with probable B-star candidates (blue crosses) from the MCPS. A 90\arcsec\ radius circle (red) is centred on the position of the SNR.}
 \label{fig8}
 \end{center}
\end{figure*}

As massive stars rarely form in isolated environments, core-collapse supernovae are most likely superposed on a stellar population rich in massive stars. We have used the Magellanic Cloud Photometric Survey \citep[MCPS]{2004AJ....128.1606Z} data to construct colour-magnitude diagrams (CMDs) and identify blue stars more massive than $\sim$8~$M_\odot$ within the area of \SNR. The CMD in Fig.~\ref{fig8} (left) indicates that most of the blue massive stars are main sequence B stars. Their spatial distribution marked in blue Fig.~\ref{fig8} (right) shows higher concentrations toward a cluster to the east and the nebulosity to the southeast of the SNR, but not within the SNR and its immediate vicinity. The progenitor of \SNR\ could be a B star (core-collapse supernova) or an accreting white dwarf in a binary system (Type~Ia supernova). To distinguish between these two possibilities, deep \emph{XMM-Newton} X-ray observations with sufficient counts for spectral analysis of plasma abundances are needed.

%%%%%%%%%%%%%%%%%%%%%%%%%%%%%%%%%%%%%%%%%%%%%%%%%%%%%%%%%%%%%%%%%%%%%%%%%%%%%%%%%%%%%%%%%%%%%%%%%%%%%%%%%%%%

\section{Conclusion}
 \label{conclusion}

We have carried out the first detailed multi-frequency study on a recently-detected LMC~\SNR, which previously had records for a pulsar at this position. We estimated a diameter of 137\arcsec$\times$128\arcsec$\pm$4\arcsec (33$\times$31~pc with 1~pc uncertainty in each direction), a spectral index ($\alpha=-0.68\pm$0.03) and a peak polarisation of $\sim$17\%$\pm$7\%. We note that even though there is no optical emission (and thus no \SII/\Halpha$>$0.4), the presence of the non-thermal radio and the X-rays do satisfy two of the three criteria for calling it an SNR. While there is a positional association between the SNR and PSR, they are not necessarily related as the ages are inconsistent. While assuming that the PSR has the same age as the SNR yields estimated kick velocities that are fairly typical, there is no radio-frequency trail of the kind seen in other SNR-PSR connections.

%%%%%%%%%%%%%%%%%%%%%%%%%%%%%%%%%%%%%%%%%%%%%%%%%%%%%%%%%%%%%%%%%%%%%%%%%%%%%%%%%%%%%%%%%%%%%%%%%%%%%%%%%%%%

\section*{Acknowledgements}
We used the {\sc karma} software package developed by the ATNF. The ATCA is part of the Australia Telescope which is funded by the Commonwealth of Australia for operation as a National Facility managed by CSIRO. We thank the MCELS team for access to the optical images. Travel to the SAAO was funded by Australian Government AINSTO AMNRF grant number 10/11-O-06. This research is supported by the Ministry of Education and Science of the Republic of Serbia through project No. 176005. We thank the referee for their excellent comments that improved this manuscript.

\bibliographystyle{mn2e}
\bibliography{MC}
\label{lastpage}
\end{document}